\documentclass[useAMS,usenatbib,onecolumn]{mnras}

\usepackage{dcolumn}
\usepackage{graphicx}
\usepackage{url}
\usepackage{color}
\usepackage{mathtools}
\usepackage{longtable}
\usepackage{booktabs}

\usepackage{multirow}
\newcommand{\cm}{cm$^{-1}$}

\newcommand{\ai}{\textit{ab initio}}

\newcommand{\Duo}{{\sc Duo}}

\newcommand{\X}{$X$~$^{2}\Pi$}
\newcommand{\Astate}{$A$~$^{2}\Sigma^{+}$}
\newcommand{\B}{$B$~$^{2}\Pi$}

\newcommand{\ket}[1]{\vert #1 \rangle  }
\newcommand{\bra}[1]{\langle #1 \vert  }

\usepackage{epstopdf}

\title[ExoMol XXXVI: SH $X$ -- $X$ and $A$ -- $X$]{ExoMol molecular line lists  XXXVI: $X$~$^{2}\Pi$ -- $X$~$^{2}\Pi$ and $A$~$^{2}\Sigma^{+}$ -- $X$~$^{2}\Pi$ transitions of SH}
\date{\today}
\author[Gorman et al.]{ Maire N. Gorman$^{1}$, Sergei N. Yurchenko$^{2}$ and Jonathan Tennyson$^{2}$\thanks{Email: j.tennyson@ucl.ac.uk}  \\
$^{1}$ Department of Physics, Aberystwyth University, Penglais, Aberystwyth, Ceredigion, UK, SY23 3BZ\\
$^{2}$ Department of Physics and Astronomy, University College London, London WC1E 6BT, UK}
\date{Accepted XXXX. Received XXXX; in original form XXXX}
\pubyear{2019}
\pagerange{\pageref{firstpage}--\pageref{lastpage}}
\begin{document}
\maketitle

%%%%%%%%%%%%%%%%%%%%%%%%%%%%%%%%%%%%%%%%%%%%%%%%%%%%%%%%%%%%%%%%%%%%%%%%%%%%%%%%%%%%%%%%%%%%%%%%%%%%%%%%%
\begin{abstract}
  The GYT line list covering rotational, rovibrational and rovibronic
  transitions of the mercapto radical SH is presented.  This work
  extends and replaces the SNaSH line list [Yurchenko et al., 2018,
  MNRAS, 478, 270] which covers the ground (electronic) \X\ state
  only. This extension is prompted by the tentative identification of
  the ultra-violet features of SH as being of importance in
  the transmission spectrum of the ultra-hot Jupiter exoplanet
  WASP-121b [Evans et al., 2018, AJ., 156, 283]. This GYT line list
  model is generated by fitting empirical potential energy, spin-orbit and electronic angular momenta
  functions to experimentally
  measured wavelengths within the \X\ and \Astate\ states and to the
  \Astate\ -- \X\ band system using \ai\
  curves as a starting reference point. The fits are compatible
  with the quoted uncertainty of the experimental data used of $\sim$
  0.03 - 0.3 \cm. The GYT line list covers wavelengths longer than 0.256
  $\mu$m and includes 7686 rovibronic states and 572~145
  transitions for $^{32}$SH.  Line lists for the $^{33}$SH, $^{34}$SH,
  $^{36}$SH and $^{32}$SD isotopologues are generated including a
  consideration of non-Born-Oppenheimer effects for SD.  The line
  lists are available from the CDS (http://cdsarc.u-strasbg.fr) and
  ExoMol (www.exomol.com) data bases.
\end{abstract}
%Exp source errors are as follows:
%Loge/Tiee(1988): 0.04 cm-1
%Ramsay 1952: sharp (0.03), diffuse (0.15)
%1939 Lewis-can't decipher - quotes to 3rd d.p in terms of Angstroms
%Johns data: 0.1 (10 band), 0.3 (20 band) for SH and 0.05 and 0.2 \cm\ for SD.
%%%%%%%%%%%%%%%%%%%%%%%%%%%%%%%%%%%%%%%%%%%%%%%%%%%%%%%%%%%%%%%%%%%%%%%%%%%%%%%%%%%%%%%%%%%%%%%%%%%%%%%%%

%%%%%%%%%%%%%%%%%%%%%%%%%%%%%%%%%%%%%%%%%%%%%%%%%%%%%%%%%%%%%%%%%%%%%%%%%%%%%%%%%%%%%%%%%%%%%%%%%%%%%%%%%
\begin{keywords}
molecular data; opacity; astronomical data bases: miscellaneous; planets and
satellites: atmospheres; stars: low-mass
\end{keywords}
%%%%%%%%%%%%%%%%%%%%%%%%%%%%%%%%%%%%%%%%%%%%%%%%%%%%%%%%%%%%%%%%%%%%%%%%%%%%%%%%%%%%%%%%%%%%%%%%%%%%%%%%%

%%%%%%%%%%%%%%%%%%%%%%%%%%%%%%%%%%%%%%%%%%%%%%%%%%%%%%%%%%%%%%%%%%%%%%%%%%%%%%%%%%%%%%%%%%%%%%%%%%%%%%%%%
\section{Introduction}
Previously, we have published line lists for the \X\ state of the main
isotopologues of the mercapto radical SH \citep{jt725}. Following
recent feedback from the exoplanet community we
extend this work to include the \Astate\ -- \X\ transitions which
feature in the UV region up to 0.256 $\mu$m.

Recently \citet{18EvSiGo.SH} used the Space Telescope Imaging
Spectrograph (STIS) instrument onboard the Hubble Space Telescope
(HST) to study both primary and secondary transits of the ultra-hot
Jupiter WASP-121b ($T_{\rm eq}$ $\geq$ 2500 K) in the near-infrared
(1.15 -1.65 $\mu$m) regions. Evans {\it et al.}
 observed a steep rise in the opacity from 0.30 -
0.47 $\mu$m in the transmission spectrum of
WASP-121b.
They   postulate that this rise
is unlikely to be the result of Rayleigh scattering by H$_{2}$ or
high-altitude aerosols as this would require an unphysically high
temperature. Instead they suggest it could be due to absorption from
the SH radical: previously \citet{09ZaMaFr.SH} used a
one-dimensional photochemical model to show that the abundances of SH could be
enhanced in hot Jupiters due to photolytic and photochemical
destruction of H$_{2}$S. WASP-121b has been studied in both
primary and secondary transits in the near-IR (1.15 -1.65 $\mu$m)
regions. \citet{18EvSiGo.SH} also note that it is important to
identify the species responsible for this ultraviolet (UV) opacity feature as it
most likely affects the global energy budget and thermal structure
%\red{and thermal inversion at lower T}
and that the complete
characterisation of this planet is an ongoing work.
%in progress as
%there is also a bump in the 1.15 -- 1.30 $\mu$m region which is not
%presently adequately explained by typical suspect sources such as TiO
%and/or VO. \red{How is relevant to SH?}

As well as exoplanets, SH is also of interest in AGB (asymptotic giant branch)
 and Mira variable
stars \citep{00YaKaRi.SH}, the Sun's atmosphere \citep{02BeLixx.SH}
and is potentially observable in brown dwarfs \citep{06ViLoFe.SH}. It
was also finally detected in the ISM (interstellar medium) by \citet{12NeFaGe.SH} following
unsuccessful searches by \citet{69MeGoLi.SH} and \citet{71HeTuxx.SH}. It was
tentatively detected in comets Halley and IRAS-Araki-Alcook
\citep{87SwWaxx.SH,88KrWaxx.SH} and more securely
subsequently \citep{92KiAhxx.SH}.
\citet{92KiAhxx.SH} used their analysis of SH in the Comet
P/Brorsen-Metcalf (1989) to infer the $g$-factors of the $A-X$ band system.
The \Astate\ -- \X\ transition of SH have also been detected in
translucent interstellar clouds by \citet{15ZhGaLi.SH} who observed
the absorption features at 3242.40 \AA\ and 3240.66 \AA\ in the
\Astate\ -- \X\ (0, 0) band. In contrast to \citet{12NeFaGe.SH}, they
found their determined abundance of SH to be in line with models for
turbulent dissipation regions (TDRs). On Earth, the oxidation of H$_{2}$S
in the troposphere produces SH which is a key species in the reactions
governing the production of acid rain \citep{94RaWiFl.SH}.

%%%%%%%%%%%%%%%%%%%%%%%
%Final SNASH paper - reworded above
%The diatomic mercapto radical SH has long been of interest to astronomers, but proved challenging to detect. It was detected definitively in the ISM \citep{12NeFaGe.SH} and AGB stars \citep{00YaKaRi.SH} and the Sun's atmosphere \citep{02BeLiXX.SH}, tentatively detected in comets \citep{87SwWaXX.SH,88KrWaXX.SH} and predicted to occur in brown dwarfs \citep{06ViLoFe.SH} and hot Jupiter exoplanets \citep{06ViLoFe.SH,09ZaMaFr.SH} as one of the major sulphur-bearing gases after H$_2$S. The ISM detection was difficult due to the location of the key rotational transition which was inaccessible both from the ground and the Herschel telescope; after a number of failed searches in the ISM \citep{69MeGoLi.SH,71HeTuxx.SH}, \citet{12NeFaGe.SH} finally detected SH in the terahertz region using SOFIA (Stratospheric Observatory For Infrared Astronomy) by its 1383 GHz $^{2}\Pi_{\frac{3}{2}}\  J=\frac{5}{2} - \frac{3}{2}$ transition.
%%%%%%%%%%%%%%%%%%%%%%%

%%%%%%%%%%This was cut out of original paper -worked into above
%The first detection of SH outside Earth was
 % by \citet{00YaKaRi.SH} who used the observational data of
  %\citet{84RiCaHa.SH} to not only detect SH in the S-type star R
 % Andromedae, a Mira variable star, but also deduced how it
 % periodically moves inward during stellar pulsations.

%In our own atmosphere, SH is known to react with NO$_{2}$, O$_{2}$ and O$_3$: SH is produced in the troposphere by oxidation of H$_2$S by the OH radical \citep{94RaWiFl.SH}.
%%%%%%%%%%%%

%%%%%%%%%
%The following was all cut out of original paper
The \Astate\ --\X\ ultra-violet (UV) absorption band considered in
this work has been detected in both the disc centre and limb of the
Sun \citep{02BeLixx.SH} in the ultraviolet regime using a combination
of experimental data of \citet{52Ramsay.SH} and \citet{95RaBeEn.SH}.
\citet{02BeLixx.SH} synthesized spectra using the model of
\citet{99GrSaxx.SH} and hence identified five unblended lines around
3300 \AA\ which they predict could be useful indicators for
determining the S abundances in G and K type stars.  However, they
noted that strong umbral lines were distorted by stray photospheric
light which prompted \citet{15Sinha.SH} to revisit the work by
applying new models and data in order to understand why SH seemed to
appear in the photosphere but not umbral regions. \citet{15Sinha.SH}
concluded that laboratory data on the $A-X$ oscillator strength is
essential to resolve the apparent paradox.
%\red{SY: It would be great resolve it now but I don't know how.}

The \Astate\ state of SH has a predissociative character due to its
$^2\Sigma^{+}$ curve crossing repulsive states of symmetry $^2\Sigma^{-}$,
$^4\Sigma^{-}$ and $^4\Pi$ \citep{97WhOrAsa.SH}, which significantly
affects the \Astate\ lifetimes. The lifetimes of this state of SH and
SD was the subject of a number theoretical and experimental studies
\citep{83TiFeWa.SH,83FrBrAn.SH,88LoTixx.SH,89KaSaIn.SH,90LoTixx.SH,90UbTexx.SH,97WhOrAs.SH,97WhOrAsa.SH,08BuMaMo.SH}.

%%%%%%%%%%%%%%%%%%%%
The spectrum of the mercapto radical SH has been studied
experimentally since the work of \citet{39GlHoxx.SH} and
\citet{39LeWhxx.SH} with over 100 experimental publications to
date. Many of these studies focus on photodissociation, and the
hyperfine and magnetically split lines of SH. The spectroscopy of the
$A$--$X$ band system was studied by \citet{39LeWhxx.SH} and
\citet{52Ramsay.SH} who undertook flash photolysis experiments using
H$_{2}$S to produce SH. \citet{39LeWhxx.SH} were able to measure the
$^{2}\Sigma^+$ $\leftarrow$ $^{2}\Pi_{3/2}$ system near 3237 \AA\ and
calculated spectroscopic constants. Then \citet{52Ramsay.SH}
measured the (0, 0) and (1, 0) bands in absorption and found the
latter to be diffuse indicating predissociation in the first
vibrational band of the \Astate\ state. Later the (2, 0) band was
photographed by \citet{61JoRaxx.SH}. The lifetime of the $\nu = 0$
\Astate\ state was measured using laser-induced fluorescence (LIF) methods and the Hanle effect by
\citet{88LoTixx.SH}. The empirical term values for the $A$ state were
obtained by \citet{90ScMeWe.SH} from the photodissociation of H$_2$S
at 121.6 nm.
%%%%%%%%%%%%%%%%%%%

%%%%%%%%%%%%%%%%%%%
At present, an experimentally limited absorption spectrum line list
has been compiled for the \Astate\ -- \X\ transition of SH by
\citet{09ZaMaFr.SH} using the RKR potential method of
\citet{73ZaScHa.SH} for the first three vibrational states of the
\Astate\ state. This method uses experimentally derived molecular
constants which are subject to errors due to i) imposing particular
Hamiltonians for limiting Hund's cases and ii) contamination from
perturbing electronic states. Here in this work, we bypass molecular
constants and instead fit potentials directly to experimentally
measured line positions and have thus computed an experimentally-tuned
theoretical line list which spans higher vibrational and rotational
levels as appropriate up to the dissociation limit of the \Astate\
state.

In this work we present an accurate and complete line list for SH
based on a mixture of \ai\ calculations and empirical refinements. This
line list supersedes the  SNaSH line list for SH \citep{jt725}, produced previously as part of the ExoMol project \citep{jt528}.
The SNaSH line list only considered transitions within the \X\ electronic ground state.

%%%%%%%%%%%%%%%%%%%
%%%%%%%%%%%%%%%%%%%%%%%%%%%%%%%%%%%%%%%%%%%%%%%%%%%%%%%%%%%%%%%%%%%%%%%%%%%%%%%%%%%%%%%%%%%%%%%%%%%%%%%%%%%%%%%%%%%%%%%%%%%%%%%%%%%%%%%%%%%%%%%%%%%%%%%%%%%%%%%%%%%%%%%%%%%%%%%%%%%%%%%%%%%%%%%%%%%%%%%%%%%%%%%%%%

%%%%%%%%%%%%%%%%%%%%%%%%%%%%%%%%%%%%%%%%%%%%%%%%%%%%%%%%%%%%%%%%%%%%%%%%%%%%%%%%%%%%%%%%%%%%%%%%%%%%%%%%%%%%%%%%%%%%%%%%%%%%%%%%%%%%%%%%%%%%%%%%%%%%%%%%%%%%%%%%%%%%%%%%%%%%%%%%%%%%%%%%%%%%%%%%%%%%%%%%%%%%%%%%%%
\section{Method}

The ExoMol methodology is well-established \citep{jt511,jt693}, so only the
key details are given below.
The construction of a line list consists of four distinct steps which are dictated by the Born-Oppenheimer approximation which decouples the Schr\"{o}dinger equation for a molecule into an electronic Schr\"{o}dinger equation and rovibronic Schr\"{o}dinger equation: the former is then used as input to the latter. These distinct steps are:
\begin{enumerate}
\item Calculation of \ai\ curves by solving the electronic Schr\"{o}dinger equation to produce potential energy curves (PECs), spin-orbit curves (SOCs), electronic angular momentum curves (EAMCs), dipole moment and transition dipole moment curves (DMCs and TDMCs);
\item Refinement of the \ai\ PECs, SOCs and EAMCs by fitting to experimental data;
\item Solving the rovibronic Schr\"{o}dinger equation using these refined curves;
\item Computing Einstein~$A$ coefficients using the eigenfunctions obtained and \ai\ DMCs and TDMCs.
\end{enumerate}

\begin{figure}
\centering
\includegraphics[width=0.80\textwidth]{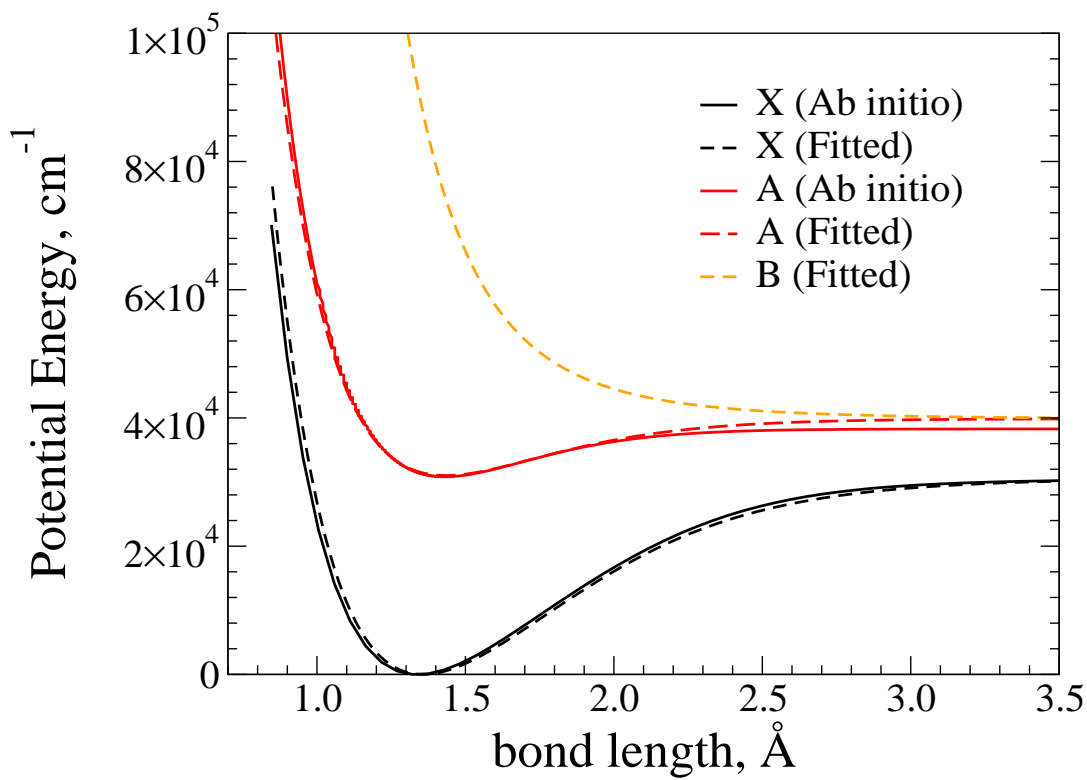}
\caption{Potential Energy Curves (PECs) used in the construction of this new extended line list.}
%Plot updated using 42 model: Friday 5th July 2019.}
\label{fig:PEC}
\end{figure}

Using MOLPRO \citep{MOLPRO},  \ai\ calculations were performed for low-lying electronic states of SH using a grid of 417 points between 0.7 and 19.2 \AA\ with more points concentrated around equilibrium. These calculations included PECs for the \Astate\ and \B\ states and the $A$--$X$ TDMC and were based on the multi-reference configuration interaction (MRCI)/aug-cc-pV5Z-DK level of theory \citep{89Dunning.ai,Woon-aug-cc-pVXZ-1993}: the initial complete active space self-consistent field (CASSCF) calculation over which the configuration interaction calculations were built was for the \X\ state only.  Figure \ref{fig:PEC} shows the selection of PEC curves used in the final line list model. The active space was selected to be (8,4,4,1) with  (3,1,1,0) closed orbitals.  The Douglass-Kroll relativistic corrections were taken into account (kroll=1).
%\red{8441/3110:  1s, 1s, 2s, 2p, 3s, 3p, 3d corresponds to 8330; 1s, 1s, 2s, 2p corresponds to 4110}.

%%%%%%%%%%%%%%%%%%%%%%%%%%%%%%%
%%%%%%%%%%%%%%%%%%%%%%%%%%%%%%%

%%%%%%%%%%%%%%%%%%%%%%%%%%%%%%%
\begin{figure}
\centering
\includegraphics[width=0.80\textwidth]{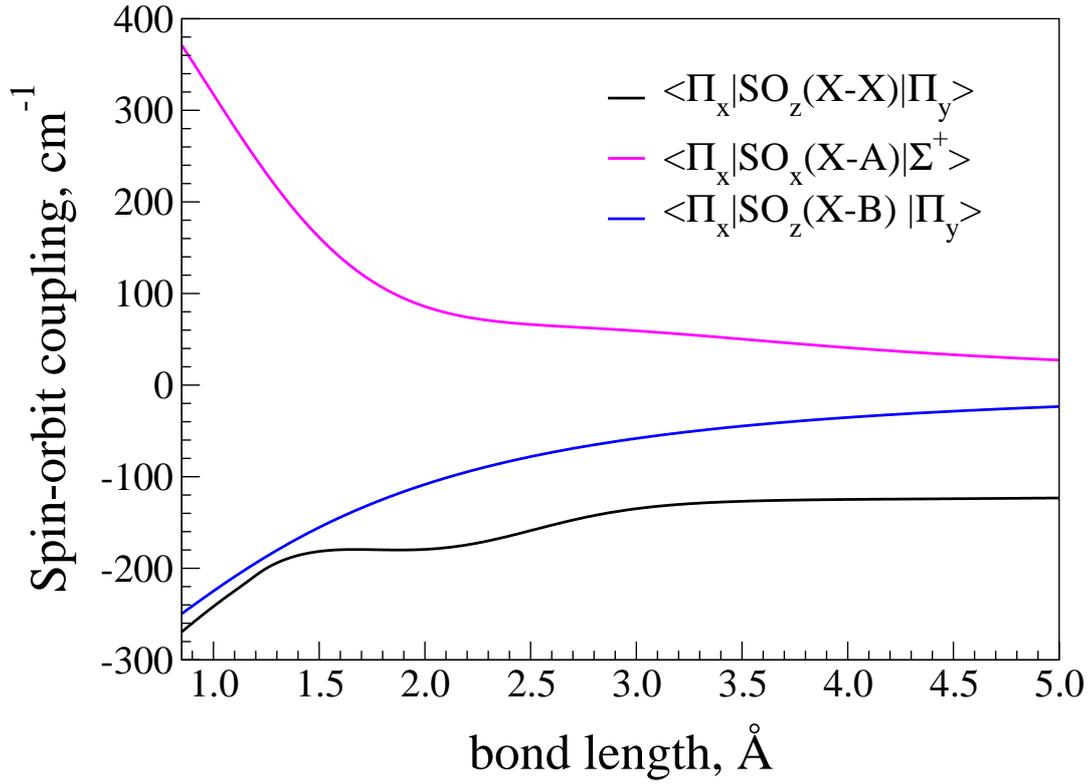}
\caption{Spin-orbit coupling curves (SOCs) included in the model.}
%Plot generated using 52 model: Thursday 30th May 2019.
\label{fig:SOC}
\end{figure}
%%%%%%%%%%%%%%%%%%%%%%%%%%%%%%%

%%%%%%%%%%%%%%%%%%%%%%%%%%%%%%%
\begin{figure}
\centering
\includegraphics[width=0.80\textwidth]{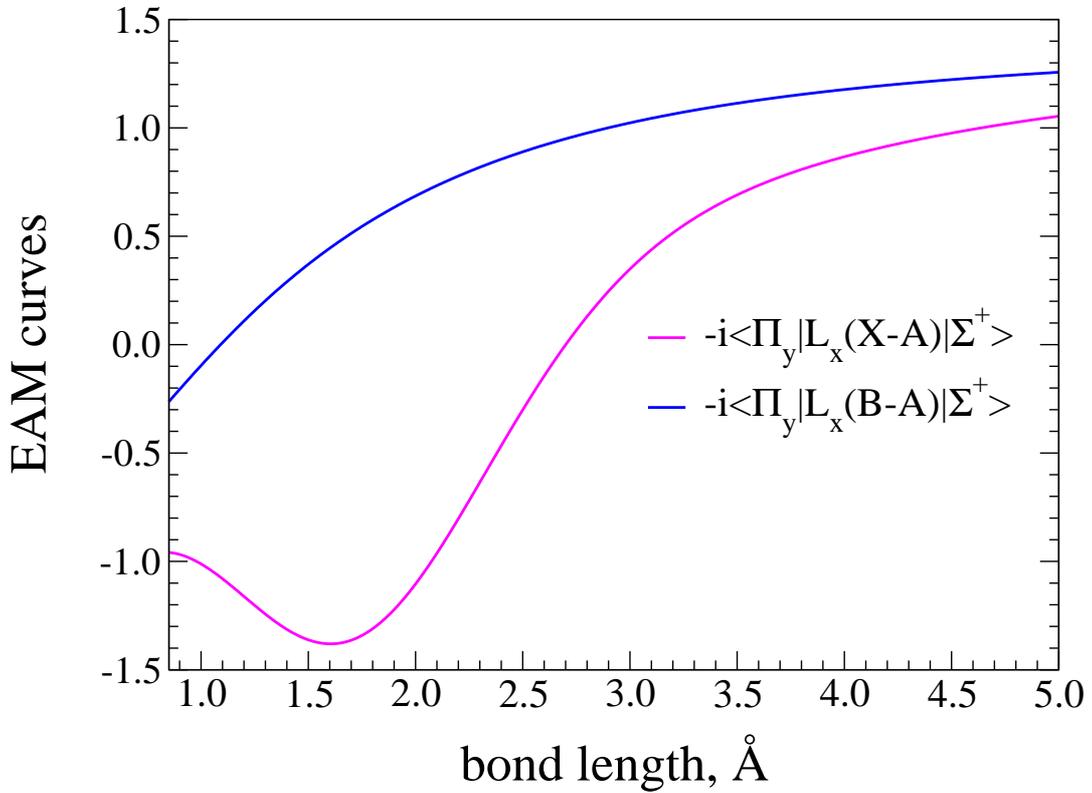}
\caption{Electronic angular momentum curves (EAMCs) included in the model.} %Plot generated using 52 model: Thursday 30th May 2019.}
\label{fig:EAMC}
\end{figure}
%%%%%%%%%%%%%%%%%%%%%%%%%%%%%%%

%%%%%%%%%%%%%%%%%%%%%%%%%%%%%%%
\begin{figure}
\centering
\includegraphics[width=0.80\textwidth]{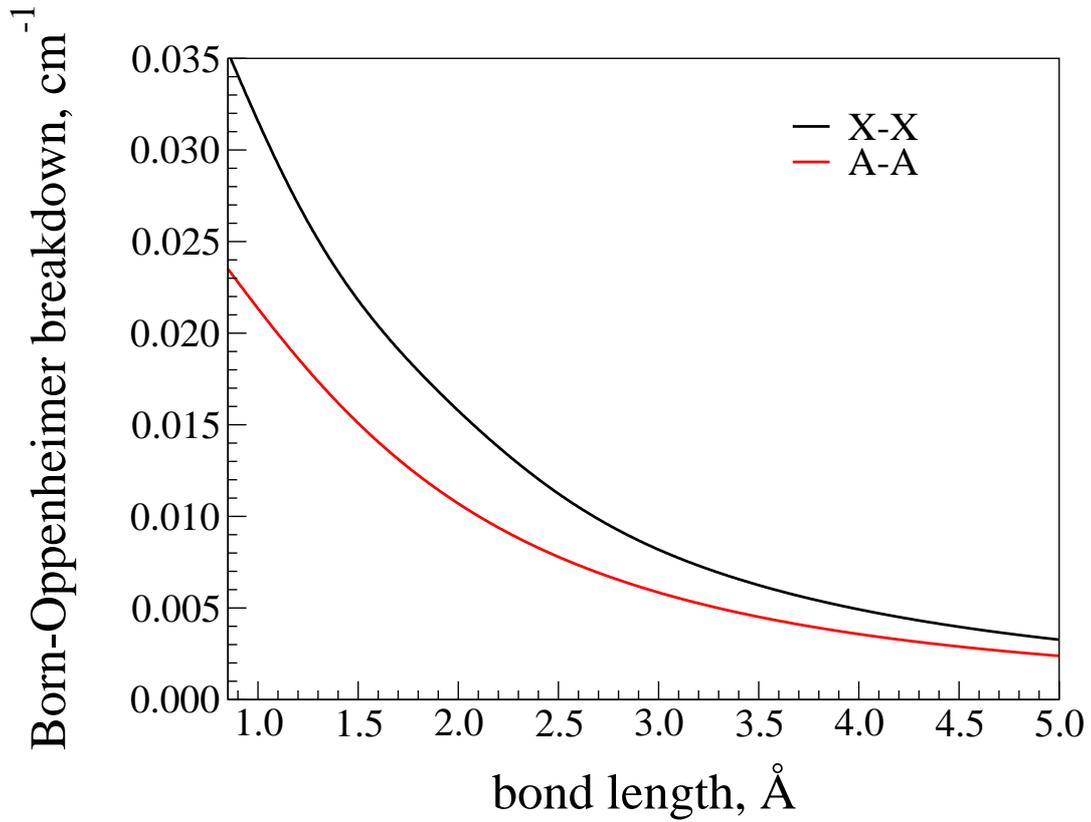}
\caption{Born-Oppenheimer breakdown curves (BOBCs) included in the model} %Plot generated using 52 model: Thursday 30th May 2019.
\label{fig:BOB}
\end{figure}
%%%%%%%%%%%%%%%%%%%%%%%%%%%%%%%

%%%%%%%%%%%%%%%%%%%%%%%%%%%%%%%
\begin{figure}
\centering
\includegraphics[width=0.80\textwidth]{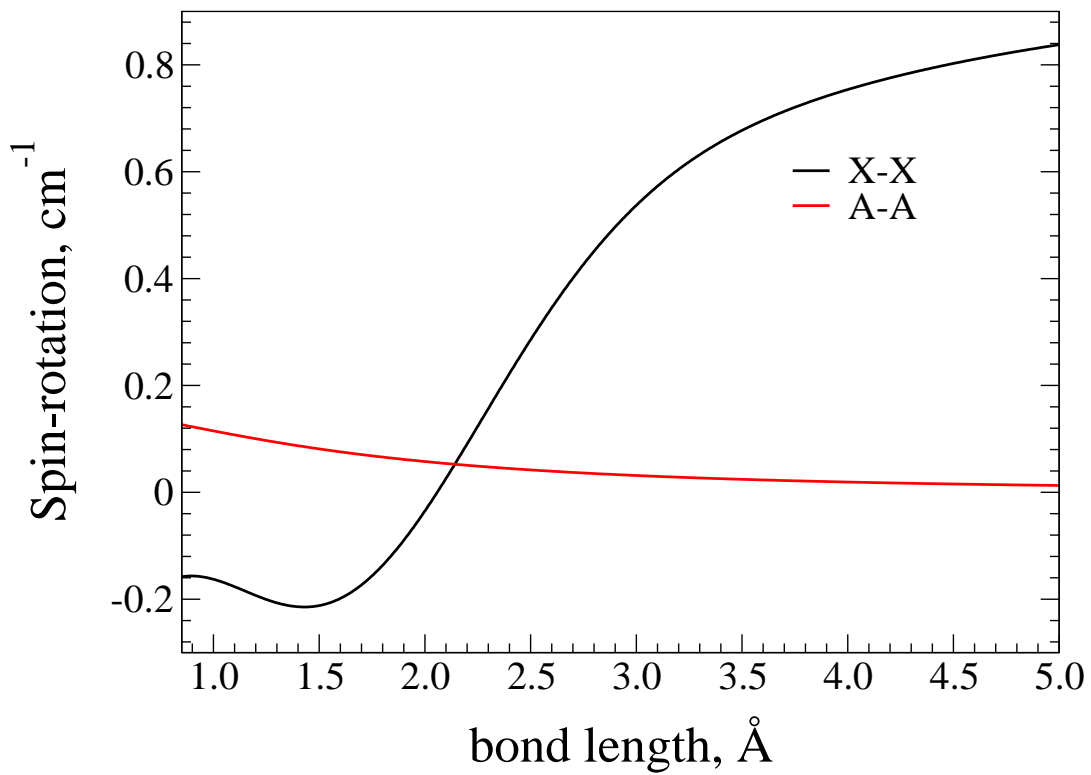}
\caption{Spin-rotation curves (SRCs) included in the model.}
%Plot updated using 52 model: Thursday 30th May 2019.
\label{fig:SR}
\end{figure}
%%%%%%%%%%%%%%%%%%%%%%%%%%%%%%%

%%%%%%%%%%%%%%%%%%%%%%%%%%%%%%%
\begin{figure}
\centering
\includegraphics[width=0.80\textwidth]{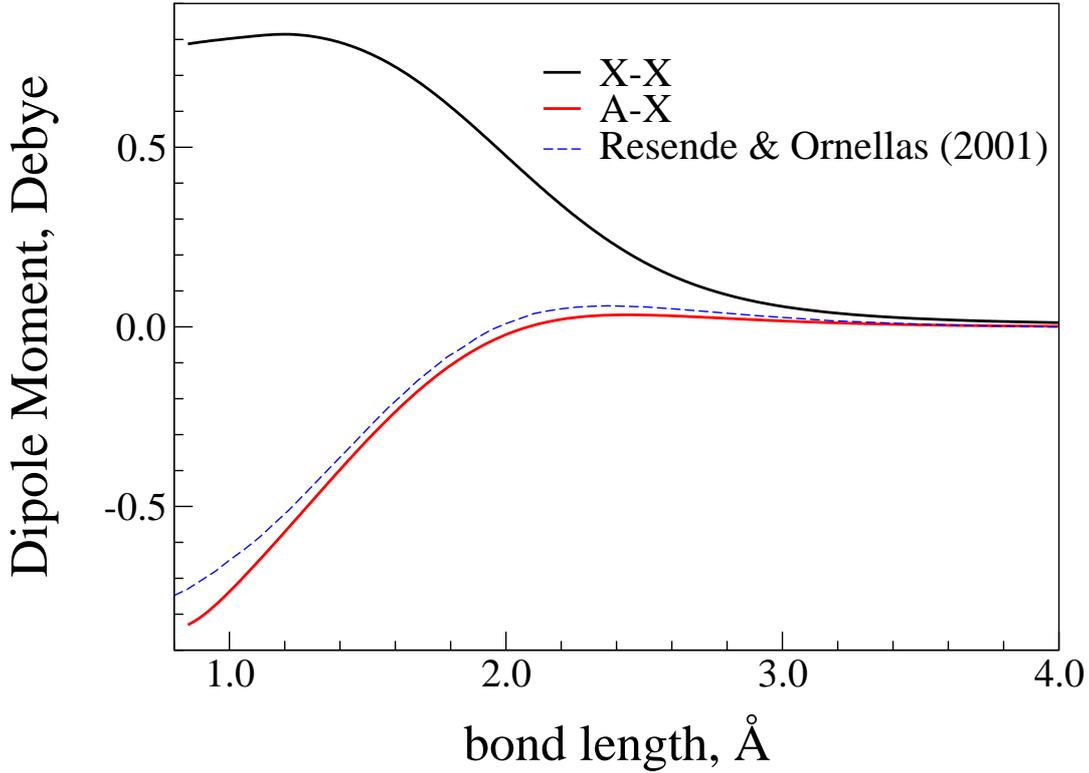}
\caption{Dipole moment  curves (DMCs/TDMCs) $X$--$X$ and $A$--$X$ included in our model. The latter is compared to an \ai\ TDMC of \protect\citet{01ReOrxxa.SH}.}
%Plot generated using 52 model: Thursday 30th May 2019.}
\label{fig:DMC}
\end{figure}
%%%%%%%%%%%%%%%%%%%%%%%%%%%%%%%
\citet{jt725} fitted the \X\ ground electronic state of SH to
available experimental data. Table \ref{t:HSlit} shows the
experimental sources used in this work to fit both the \X\ and
\Astate\ electronic states.
%%%%%%%%%%%%%%%%%%%%%%%%%%%%
\begin{table}
\centering
\caption{List of experimental data used in refinement of the SH (and SD) \X\ and \Astate\ potential energy curves.}
\begin{tabular}{lrccr}
\hline\hline
Source			& No. of transitions & El. Band &Vib. bands		& $J_{\rm max}$ \\ \hline
\citet{83BeAmWo.SH}	&	50		&$X$& 	(1-0), $X$			&	11.5	\\
\citet{84WiDaxx.SH}	&	285		&$X$&	(1-0), (2-1), (3-2)		&	34.5	\\
\citet{95RaBeEn.SH}	&	175		&$X$&	(1-0), (2-1), (3-2), (4-3)	&	16.5	\\
\citet{00YaKaRi.SH}	&	30		&$X$&	(1-0), (2-1), (3-2)		&	25.5	\\
\citet{11ElMaGu.SH}	&	6		&$X$&	(0-0)				&	4.5	\\
\citet{12MaElPi.SH} &	70		&$X$&	(0-0), (1-1)				&	16.5	\\ \hline
%%%%%%%%%%%%%%%%%%%%%%%%%%%%
\citet{39LeWhxx.SH}	&	45		&$A$&	(0-0)				&	9.5	\\
\citet{52Ramsay.SH}	&	170		&$A$&	(0-0), (1-0)			&	12.5	\\
\citet{61JoRaxx.SH}	&	146		&$A$&	(1-0), (2-0)			&	9.5	\\
\citet{88LoTixx.SH}	&	12		&$A$&	(0-0)				&	6.5	\\ \hline
%%%%%%%%%%%%%%%%%%%%%%%%%%%%
\citet{52Ramsay.SH}	& 189			& $A$ (SD)	& (0-0), (1-0)			& 15.5		\\
\citet{69PaPaxx.SH}	& 21			& $A$ (SD)	& (0-1)				& 16.5		\\
\citet{61JoRaxx.SH}	& 242			& $A$ (SD)	 & (1-0), (2-0)			& 15.5		\\ \hline \hline
%%%%%%%%%%%%%%%%%%%%%%%%%%%%
\end{tabular}
\label{t:HSlit}
\end{table}
%83UbTeDy.SH.pdf: Hyperfine and $\rho$ splittings
%88LoTixx.SH.pdf: A-X spectra, lifetime measurement, improved line position
%90ScMeWe.SH.pdf: ~64 A state transitions: omega, omega-xhi, Re, De, alpha parameters determined
%93MoLaMo.SH.pdf: SD data
%%%%%%%%%%%%%%%%%%%%%%%%%%%%%%%%%%%%%%%%%%%%%%%%%%%%%%%%

Despite the experimental interest of SH, there is a dearth of
experimentally measured rovibronic transitions required for refinement
for the for \Astate\ -- \X\ system. As for any electronic transition
system, the accuracy of a line list model generated is dependent upon
the (i) vibrational and rotational coverage of available
experimentally measured line positions and (ii) the measurement
accuracy which is experimentally possible for the wavenumber range
which the system is located. Here we are unfortunately limited to four
experimental sources
\citep{39LeWhxx.SH,52Ramsay.SH,61JoRaxx.SH,88LoTixx.SH} which only span
the (0, 0), (1, 0), (2,0) $A$-$X$ vibronic bands with rotational
coverage up to $J$ = 12.5. These bands occur at $\sim$ 31~000 \cm\
(0.32 $\mu$m) and have a quoted experimentally measured accuracy of
between $\approx$ 0.1 - 0.5 \cm. We do not use the work of
\citet{90ScMeWe.SH} which  contains empirical vibronic
\Astate\ term values ($v=0\ldots 4$, $N=0\ldots 40$) as these are  not
sufficiently accurate for the present study: \citeauthor{90ScMeWe.SH}
compared their term values to those determined using theoretical
extended Rydberg and Morse potentials and calculated discrepancies of
$\sim$ 1 -- 66 \cm. More pertinently, these term values were
obtained via fitting spectroscopic constants to observed spectra which
hence introduces potential error due to the effect of nearby coupling
states. %comment - mng2 - checked paper - really was this bad....
%%%%%%%%%%%%%%%%%%%%

%%%%%%%%%%%%%%%%%%%%
In order to fit the \Astate\ state PEC, the experimental frequencies were collected from \citet{39LeWhxx.SH,52Ramsay.SH,61JoRaxx.SH} and \citet{88LoTixx.SH}, all representing the $A$--$X$ transitions, the (0,0), (1,0) and (2,0) vibronic bands (see Table~\ref{t:HSlit}).
Including the interactions with other nearby lying electronic states is crucial to achieving good reproduction of the experimental frequencies. To this end the SO and  EAM couplings with the \X\  state were introduced and varied. The latter affected the quality of the $X$ energies, which in turn required us to refine our \X\ state spectroscopic model \citep{jt725} in a global fit to the experimental data listed in Table~\ref{t:HSlit}. For any refinement, where possible, it is imperative to use measured transitions as opposed to experimentally determined spectroscopic constants in order to account for perturbations due to higher lying electronic states which are often not explicitly accounted for during the processes of obtaining spectroscopic constants for individual electronic states.
The \X\ state, however, did not suffice to account for the spin splitting of the \Astate, and another $^2\Pi$ state, \B, was added. The \B\ state has a repulsive PEC and is much closer to \Astate,  see Fig.~\ref{fig:PEC}. Our final spectroscopic model thus consists of the following components:
\begin{itemize}
  \item Three fitted PECs, \X, \Astate\ and \B, Fig.~\ref{fig:PEC};
  \item Three fitted SOCs, $X$--$X$, $X$--$A$, $A$--$B$, Fig.~\ref{fig:SOC};
  \item Two fitted EAMc, $X$--$A$ and $A$--$B$, Fig.~\ref{fig:EAMC};
  \item Two fitted Born-Oppenheimer Breakdown curves (BOBC), $X$ and $A$, Fig.~\ref{fig:BOB};
  \item Two fitted spin-rotation curves (SRC), $X$ and $A$, Fig.~\ref{fig:SR};
  \item Two \ai\ dipole moment curves (DMC), a diagonal $X$--$X$ and transition $A$--$X$, Fig.~\ref{fig:DMC}.
\end{itemize}

Figure~\ref{fig:DMC} compares our \ai\ TDMC with that from a similar \ai\ study computed by \citet{01ReOrxxa.SH} using the MRCI/aug-cc-pV5Z level of theory, which shows generally good agreement.

%%%%%%%%%%%%%%%%%%%%%%%%%%%%%%
%Not sure what reference is best for DVR so put reference to Duo program.

\begin{figure}
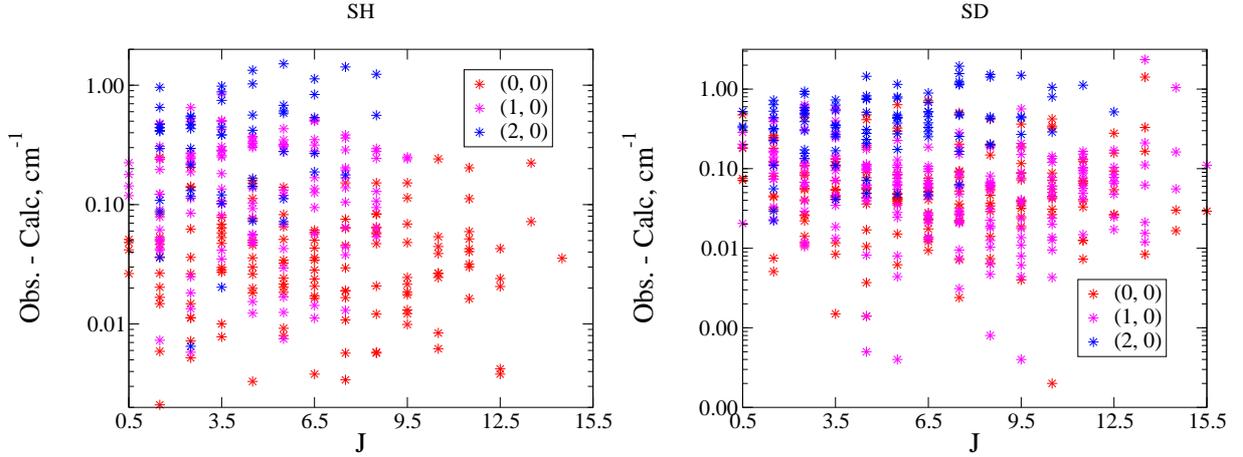

\centering
\includegraphics[width=0.45\textwidth]{OC-M54-V1.eps}
\includegraphics[width=0.45\textwidth]{OC-SD-V1.eps}
\caption{Obs. - Calc. residuals for transitions from the three vibronic bands \Astate\ -- \X\ of SH (left) and SD (right). }
%Updated using model 54, July 5th
\label{fig:OC-AX}
\end{figure}

With this model the rovibronic Schr\"{o}dinger equation was then
solved using the program \Duo\ \citep{jt609}. \Duo\ is the
custom-built program developed within the ExoMol group for calculating
line lists for general, open-shell diatomics represented by
arbitrary number of couplings \citep{jt632}. The vibrational basis set comprised 120
functions, 40+40+40 for each of the three electronic states $X$, $A$ and $B$,
obtained by solving these three independent problems with the Sinc
discrete variable representation (DVR)
on a grid of 501 points between 0.85 and 5.00 \AA\
\citep{jt609}. Note that although the \B\ state is dissociative, since \Duo\
can only deal with bound states basis sets, by the nature of Sinc DVR,
the corresponding PEC had an effective infinite wall at the right end
of the grid. Using this model we were able to achieve the accuracy of
the fit comparable with the quality of the corresponding experimental
data, $\sim$~0.001--0.01~\cm\ for the $X$--$X$ line positions
 and $\sim$0.05--0.5~\cm\ for the $A$--$X$ frequencies,
see Table~\ref{t:obcalcSH} and Figure~\ref{fig:OC-AX}. The final
root-mean-squares (rms) errors for the $X$--$X$ and $A$--$X$
transition wavenumber frequencies are 0.06 and 0.3~\cm, respectively.
%MNG2: Taken out file -done in excel: model-52: Wed 5th July.
%%%%%%%%%%%%%%%%%%%%%%%%%%%%%%%
%%%%%%%%%%%%%%%%%%%%%%%%%%%%%%%
Following our previous work on SH \citep{jt725}, both the \X\ and \Astate\ PECs were represented using an Extended Morse Oscillator (EMO) function \citep{EMO} as given by
\begin{equation}\label{e:EMO}
V(r)=V_{\rm e}\;\;+\;\;(A_{\rm e} - V_{\rm
e})\left[1\;\;-\;\;\exp\left(-\sum_{k=0}^{N} B_{k}\xi_p^{k}(r-r_{\rm e})
\right)\right]^2,
\end{equation}
where $A_{\rm e} - V_{\rm e} = D_{\rm e}$ is the dissociation energy, $A_{\rm e}$ is the corresponding asymptote, $r_{\rm e}$ is an equilibrium distance of the PEC, and $\xi_p$ is the \v{S}urkus variable \citep{84SuRaBo.method} given by
\begin{equation}
\label{e:surkus:2}
\xi_p= \frac{r^{p}-r^{p}_{\rm e}}{r^{p}+r^{p}_{\rm e }}
\end{equation}
with $V_{\rm e} = 0$ for the \X\ state.
The dissociation energy of the $D_{\rm e}^{(X)}$ of the \X\ state was fixed to the
\ai\ value recommended by  \citet{03CsLeBu.SH} of  3.791~eV ($D_{0}^{(X)} = 3.625$~eV), which agrees well with the experimental value
$D_{0}^{(X)} = $ 3.62 $\pm$ 0.03 eV of \citet{91CoBaLe.SH}. The asymptote limit of the \Astate\ state was estimated as and fixed to
$D_{\rm e}^{(X)}+\Delta E(S^{^1{D}}) = 4.94$~eV  with the atomic excitation energy of sulphur $\Delta E^{S({^1D})} = 1.146$~eV as taken from NIST \citep{NIST}.

The repulsive \B\ PEC was represented using the following hyperbolic form:
$$
V(r)=A_{\rm e} + \frac{B_6}{r^6},
$$
where the asymptote $A_{\rm e}$ of the \B\ state was fixed with the asymptote of  the \Astate\ state, $A_{\rm e} = 4.94$~eV. %Checked - mng2-Sun 8th June - in linelist-52-modle. %C6 has value in Linelist-52 model of  2.98088692713112e+05, Ae is of value 45238.6092

Different couplings and corrections  between different states were modelled using either the expansion:
\begin{equation}
\label{e:bob}
F(r)=\sum^{N}_{k=0}B_{k}\, z^{k} (1-\xi_p) + \xi_p\, B_{\infty},
\end{equation}
where $z$ is either taken as the \v{S}urkus variable $z=\xi_p$ (for $A$--$X$ EAMC and $X$, $A$ BOBCs)  or using the
damped-coordinate polynomial given by (for SOCs, EAMCs $A$--$B$, SRCs):
\begin{equation}\label{e:damp}
z = (r-r_{\rm ref})\, e^{-\beta_2 (r-r_{\rm ref})^2-\beta_4 (r - r_{\rm ref})^4},
\end{equation}
see also \citet{jt703} and \citet{jt711}.  Here $r_{\rm ref}$ is a
reference position chosen to be close to $r_{\rm e}$ of \X\ and $\beta_2$ and
$\beta_4$ are damping factors. In contrast
to line positions, which are greatly improved by refining PECs using
experimentally measured line positions, this is seldom true for the line intensities, for which DM and
TDM curves are usually best computed \ai\ \citep{jt573}. The line
intensities, which are directly based on the \ai\ dipole moment
curves, often suffer from the numerical noise, especially those from
the overtone transitions \citep{16MeMeSt}. In order to reduce this noise it was sufficient to apply a dipole moment cutoff of $10^{-8}$~D to the vibrational transition matrix elements, see, e.g., \citet{jt732}. When computing these matrix elements, the original \ai\ $X$--$X$ and $A$--$X$ (T)DMCs were mapped on the \Duo\ grid using the standard cubic spline interpolation technique.
The final spectroscopic model in the form of the \Duo\ input file is provided as part of the supplementary data to this paper and can be also found at \href{www.exomol.com}{www.exomol.com}. This also includes our fitting set of experimental frequencies.

%%%%%%%%%%%%%%%%%%%

%%%%%%%%%%%%%%%%%%%%%%%%%%%%%%
%%%%%%%%%%%%%%%%%%%%%%%%%%%%%%%%%%%%%%%%%%%%%%%%%%%%%%%%%%%%%%%%%%%%%%%%%%%%%%%%%%%%%%%%%%%%%%%%%%%%%%%%
A sample of the refinement fit to the data of \citet{52Ramsay.SH} for
the \X\ -- \Astate\ system is shown in Table~\ref{t:obcalcSH}. The
Obs.--Calc. values of the EMO fit are comparable to the actual
experimental measurement uncertainty. Although these may seem from the
outset to be of much lower accuracy compared to previous ExoMol line
lists for non-transition metal containing diatomics, given the regime
of $\approx$ 30~000 \cm\, an uncertainty of 0.5 \cm\ corresponds to a
resolution $R$ of $\approx$ 60~000 which is within the realms of high
spectral resolution methods of characterising exoplanet atmospheres
\citep{04Snellen,14DeBiBr}.
%Previous comments: Our results differ significantly from 90ScMeWe for v=2,3,4,5. Sc paper had errors between theory and experiment of up to 66 \cm.

For all isotopologues of SH-type ($^{33}$SH, $^{34}$SH, $^{36}$SH) we
use the same empirical model (PECs, SOCs, EAMCs, SRCs and BOBCs)
developed for $^{32}$SH. When solving the Schr\"{o}dinger equation,
the mass of sulphur is replaced by the nuclear mass of
the corresponding isotope. For the isotopologue $^{32}$SD, however,
this approach usually leads to too large discrepancies with the
experimental line positions. We have therefore refined the $^{32}$SH
model by fitting to the experimental line positions of $^{32}$SD from
the $A$--$X$ band, which were taken from the same sources listed in
Table~\ref{t:HSlit}. Ideally, this refinement should be only applied to the BOB-correction. However, due to the lack of the experimental data on the \X\ state for SD, we could not build an accurate and self-consistent model without including other curves into the fit. We have therefore
varied the parameters of  PEC($A$), PEC ($B$), BOBC ($X$), BOBC ($A$), SRC ($X$) and SRC ($A$) by fitting to the SD line positions from \citet{52Ramsay.SH,61JoRaxx.SH} and \citet{69PaPaxx.SH}. All other curves were fixed to the those from the spectroscopic model of $^{32}$SH, including PEC($X$), SOC ($X$), SOC($X$--$A$), SOC($X$--$B$), SOC($A$--$B$), EAMC($X$--$A$), EAMC($A$--$B$). The rms error of this fit to the $A$--$X$ line positions if SD is 0.4~\cm. The spectroscopic model for SD is included into the supplementary material as a \Duo\ input file.
%%%%%%%%%%%%%%%%%%%%%%%%%%%%%%%%%%%%%%%%%%%%%%%%%%%%%%%%%%%%%%%%%%%%%%%%%%%%%%%%%%%%%%%%%%%%%%%%%%%%%%%%%%%%%%%%%%%%%%%%%%%%%%%%%%%%%%%%%%%%

%%%%%%%%%%%%%%%%%%%%%%%%%%%%%%%%%%%%%%%%%%%%%%%%%%%%%%%%%%%%%%%%%%%%%%%%%%%%%%%%%%%%%%%%%%%%%%%%%%%%%%%%%%%%%%%%%%%%%%%%%%%%%%%%%%%%%%%%%%%%
\begin{table}
\centering
\caption{Example of Obs.–Calc. residuals, in \cm\ for the \Astate\--\X\ transitions of SH. Here $J'$ = $J''$+1 with the quantum labels denoting those of the \X\ state.}
%Updated Friday 5th July, Model 54
\begin{tabular}{crccrrr}
\hline
\hline
Vibrational band & $J$ & $+/-$  & \multicolumn{1}{c}{$\Omega$} &  \multicolumn{1}{c}{Obs.} & \multicolumn{1}{c}{Calc.} & \multicolumn{1}{c}{Obs.-Calc.} \\
\hline
(0, 0)	&	0.5	&	+	&	0.5	&	30481.38	&	30481.34	&	0.04	\\
(0, 0)	&	1.5	&	+	&	1.5	&	30931.39	&	30931.36	&	0.03	\\
(0, 0)	&	2.5	&	-	&	-1.5	&	30951.16	&	30951.15	&	0.01	\\
(0, 0)	&	2.5	&	+	&	0.5	&	30487.22	&	30487.21	&	0.01	\\
(0, 0)	&	3.5	&	+	&	0.5	&	30565.76	&	30565.68	&	0.08	\\
(0, 0)	&	4.5	&	-	&	-1.5	&	30984.57	&	30984.60	&	-0.03	\\
(0, 0)	&	4.5	&	+	&	0.5	&	30481.75	&	30481.70	&	0.05	\\
(0, 0)	&	5.5	&	+	&	0.5	&	30585.97	&	30585.90	&	0.07	\\
(0, 0)	&	6.5	&	+	&	1.5	&	30881.06	&	30881.02	&	0.04	\\
(0, 0)	&	7.5	&	+	&	0.5	&	30594.27	&	30594.25	&	0.02	\\
(0, 0)	&	8.5	&	+	&	0.5	&	30436.10	&	30436.15	&	-0.05	\\
(0, 0)	&	9.5	&	-	&	-1.5	&	30854.66	&	30854.65	&	0.01	\\
(0, 0)	&	9.5	&	+	&	0.5	&	30590.45	&	30590.47	&	-0.02	\\
(0, 0)	&	11.5	&	+	&	1.5	&	31030.03	&	31030.07	&	-0.04	\\
(0, 0)	&	12.5	&	+	&	1.5	&	30807.05	&	30807.03	&	0.02	\\
(1, 0)	&	1.5	&	-	&	-1.5	&	32664.21	&	32664.13	&	0.08	\\
(1, 0)	&	1.5	&	-	&	-1.5	&	32664.18	&	32664.13	&	0.05	\\
(1, 0)	&	2.5	&	-	&	-0.5	&	32325.46	&	32325.68	&	-0.22	\\
(1, 0)	&	3.5	&	+	&	1.5	&	32738.06	&	32738.34	&	-0.28	\\
(1, 0)	&	4.5	&	+	&	0.5	&	32250.82	&	32250.81	&	0.01	\\
(1, 0)	&	4.5	&	+	&	1.5	&	32656.80	&	32656.88	&	-0.08	\\
(1, 0)	&	5.5	&	-	&	-1.5	&	32648.45	&	32648.38	&	0.07	\\
(1, 0)	&	6.5	&	+	&	1.5	&	32636.85	&	32636.76	&	0.09	\\
(1, 0)	&	7.5	&	-	&	-1.5	&	32622.07	&	32621.93	&	0.14	\\
(1, 0)	&	7.5	&	-	&	-0.5	&	32199.32	&	32199.36	&	-0.04	\\
(2, 0)	&	1.5	&	+	&	1.5	&	34293.20	&	34292.73	&	0.47	\\
(2, 0)	&	2.5	&	-	&	-1.5	&	34304.50	&	34304.04	&	0.46	\\
(2, 0)	&	2.5	&	+	&	0.5	&	33848.00	&	33848.22	&	-0.22	\\
(2, 0)	&	4.5	&	+	&	0.5	&	33823.20	&	33823.37	&	-0.17	\\
(2, 0)	&	5.5	&	-	&	-1.5	&	34213.90	&	34214.18	&	-0.28	\\
(2, 0)	&	5.5	&	-	&	-0.5	&	33803.50	&	33803.39	&	0.11	\\
(2, 0)	&	6.5	&	+	&	1.5	&	34194.10	&	34194.64	&	-0.54	\\
\hline\hline
\end{tabular}
\label{t:obcalcSH}
\end{table}
%%%%%%%%%%%%%%%%%%%%%%%%%%%%%%%%%%%%%%%%%%%%%%%%%%%%%%%%%%%%%%%%%%%%%%%%%%%%%%%%%%%%%%%%%%%%%%%%%%%%%%%%%%%%%%%%%%%%%%%%%%%%%%%%%%%%%%%%%%%

\section{Line list}
%LL stats: SH-52 model, Thu 30th May.

\begin{figure}
\centering
\includegraphics[width=0.80\textwidth]{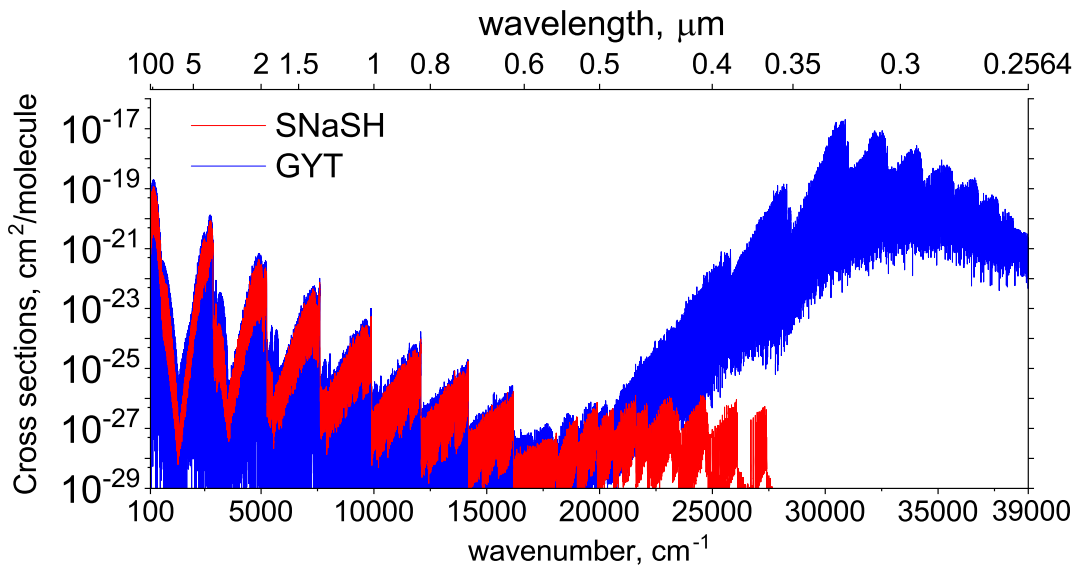} %Model 52, Enermax 50, 000
\caption{Simulated $^{32}$SH absorption spectra at 1500 K for SNaSH \citep{jt725} and the new GYT line lists. A Gaussian profile of half width of half maximum (HWHM) of 1~\cm\ was used.} %replaced reference -->> previously pointing to ORBYTS not SNASH paper.
\label{fig:COMP}
\end{figure}

\begin{figure}
\centering
\includegraphics[width=0.80\textwidth]{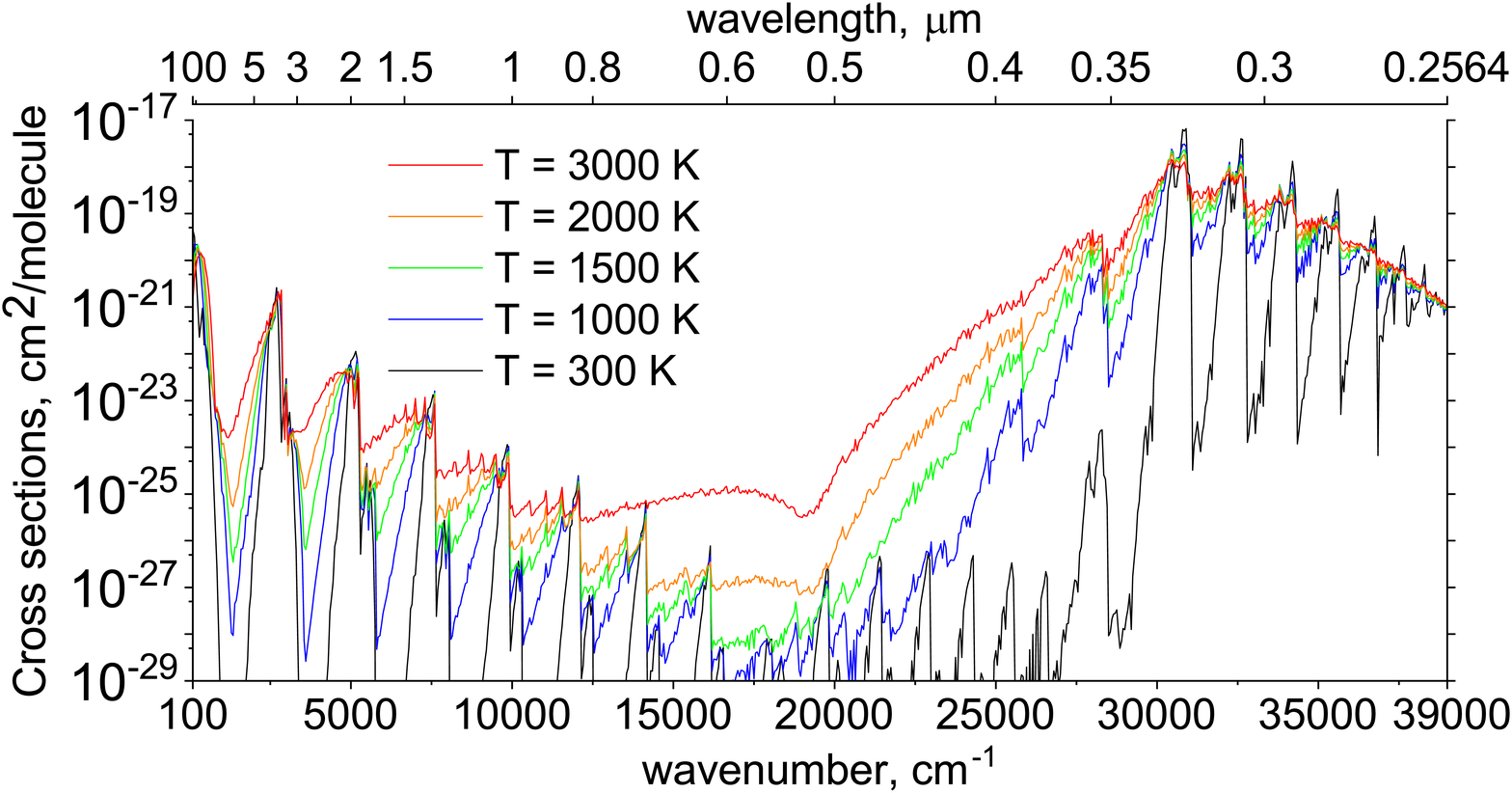} %Model 52, Enermax 50, 000
\caption{Simulated absorption spectra of $^{32}$SH for a range of temperatures; the curves become systematically
flatter as $T$ increases. A Gaussian profile of half width of half maximum (HWHM) of 1~\cm\ was used.}
\label{fig:temp}
\end{figure}

Using the final spectroscopic model in \Duo, line lists (called GYT) for 5
isotopologues of SH were computed, $^{32}$SH, $^{33}$SH, $^{34}$SH,
$^{36}$SH and $^{32}$SD. The upper states were truncated at the asymptote $A_{\rm 0}$  of the \Astate\ ($\sim$ 39~000~\cm), while the lower states were limited by the dissociation energy of the \X\ state $D_0$ ($\sim$ 31~000~\cm). Thus the GYT line lists covers the wavenumber range up to 39~000~\cm\ ($>$ 0.256 $\mu$m). The $^{32}$SH line list contains 7686 $X$ and
$A$ rovibronic states and 572~145 transitions, covering both
bands $X$--$X$ and $A$--$X$. Our new GYT line lists supersede the
SNaSH line lists for SH \citep{jt725} and extends coverage into the UV
regime. Figure \ref{fig:COMP} shows a comparison of the new GYT  line list with
SNaSH. Below 10~000 \cm\ the two line lists agree well; above this
there are difference due to the inclusion of the \Astate. Besides
the strong UV absorption of the \Astate\ there are differences
in the \X\ state absorption in the visible. Table
\ref{t:stats} showcases the size of the two line lists for
the various main isototpologues considered. Figure~\ref{fig:temp}
shows the temperature dependence of the SH spectra simulated using the \textsc{ExoCross} program \citep{jt708}.
Although the partition functions for the SH isotopologues were updated to include the $A$ state energies, they do not differ significantly from our previous partition functions due to very high energy excitations of those $A$ states.

%%%%%%%%%%%%%%%%%%%%%%%%%%%%%%%%%%%%%%%%%%%%%%%%%%%%%%%%%%%%%%%%%%%%%
%%%%%%%%%%%%%%%%%%%%%%%%%%%%%%%%%%%%%%%%%%%%%%%%%%%%%%%%%%%%%%%%%%%%%
%%%%%%%%%%%%%%%%%%%%%%%%%%%%%%%%%%%%%%%%%%%%%%%%%%%%%%%%%%%%%%%%%%%%%

\begin{figure}
\centering
\includegraphics[width=0.45\textwidth]{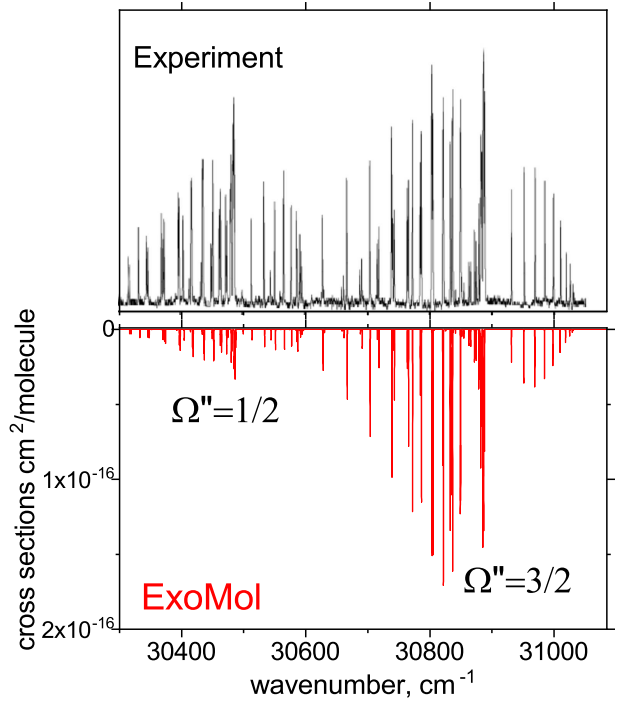}
\caption{Comparison of the absorption spectra of SH at $T=1800$~K computed using our line list (Gaussian line profile with HWHM of 0.1~\cm) with a laser-induced dispersed fluorescence  spectrum of  $A$~$^{2}\Sigma^{+}$ -- $X$~$^{2}\Pi_{3/2}$  by  \citet{10TsLixx.SH} for the (0, 0) band. }
\label{fig:10TsLixx}
\end{figure}

\begin{figure}
\centering
\includegraphics[width=0.80\textwidth]{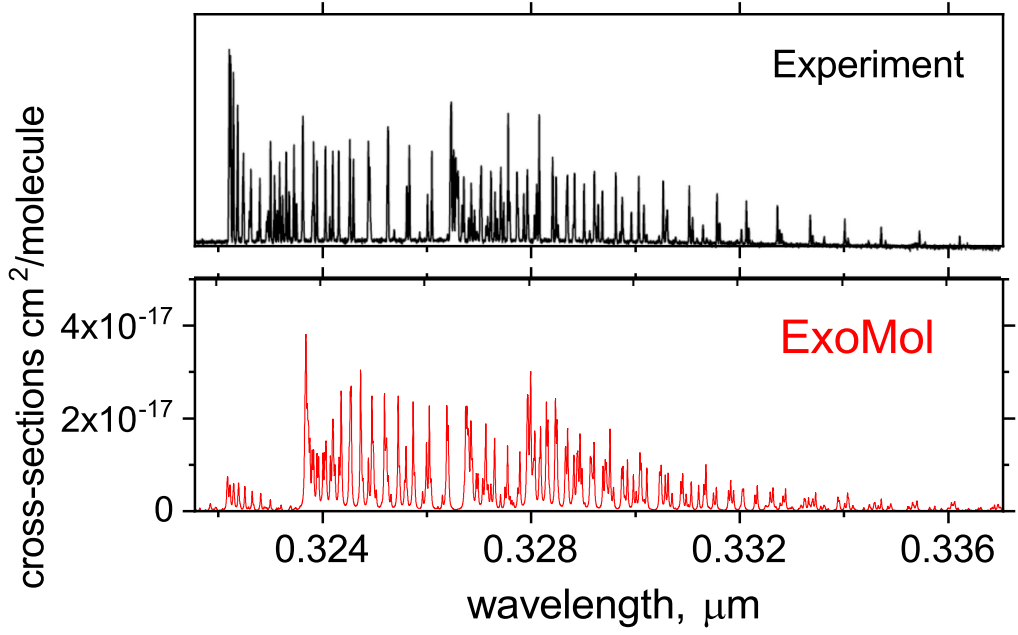}
\caption{Comparison of the new line list with the experimentally measured spectrum of the (0, 0) $A$--$X$ band by \citet{08BuMaMo.SH} using the cavity ringdown spectroscopy (CRDS). Considering a good agreement of our fit for the (0,0) $A$--$X$ band (see  Fig.~\protect\ref{fig:10TsLixx})  we believe that the shift of $\sim$0.014~$\mu$m can be attributed to a calibration problem in the experimental data.}
\label{fig:08BuMaMo.comp}
\end{figure}

\begin{figure}
\centering
\includegraphics[width=0.80\textwidth]{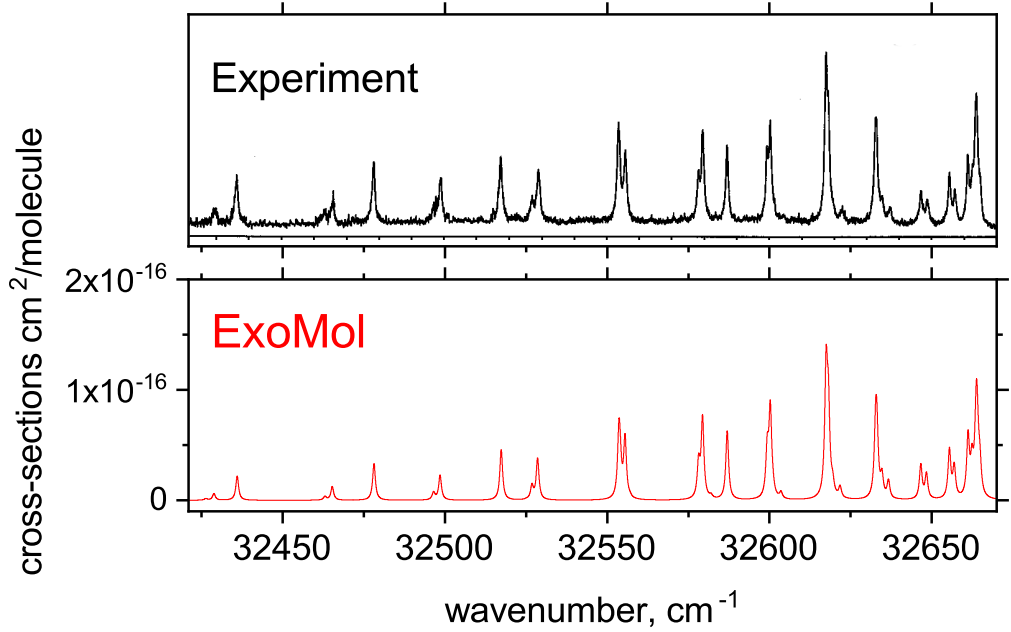}
\caption{Comparison of the absorption spectra ($T=300$~K) computed using our line list with the cavity ringdown spectroscopy (CRDS) spectrum recorded by \citet{97WhOrAsa.SH} for the (1, 0) $A$--$X$ band. A Voigt profile with $\gamma = 0.5$~\cm\ was used. }
\label{fig:97WhOrA1-comp}
\end{figure}

\begin{figure}
\centering
\includegraphics[width=0.80\textwidth]{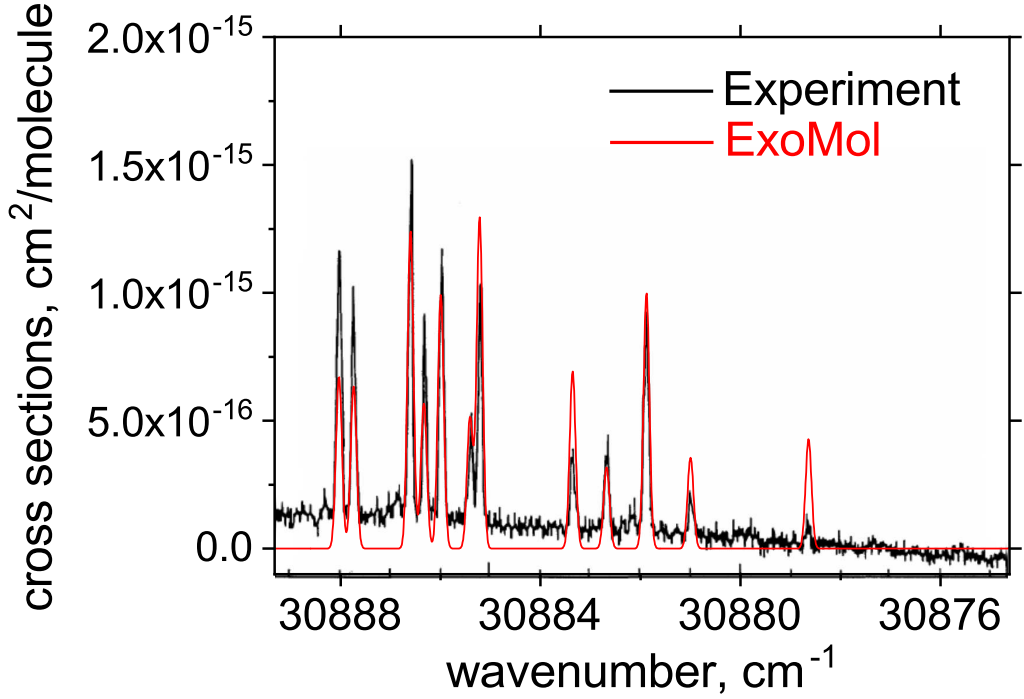}
\caption{Comparison of the absorption spectra of SH at $T=300$~K computed using our line list  with the LIF excitation spectrum of the $A$~$^{2}\Sigma^{+}$ -- $X$~$^{2}\Pi_{3/2}$  by  \citet{88LoTixx.SH} for the (0, 0) $A$--$X$ band. A Gaussian line profile with the half-width-half-maximum (HWHM) of 0.07~\cm\ was used. }
\label{fig:88LoTixx}
\end{figure}

\begin{figure}
\centering
\includegraphics[width=0.80\textwidth]{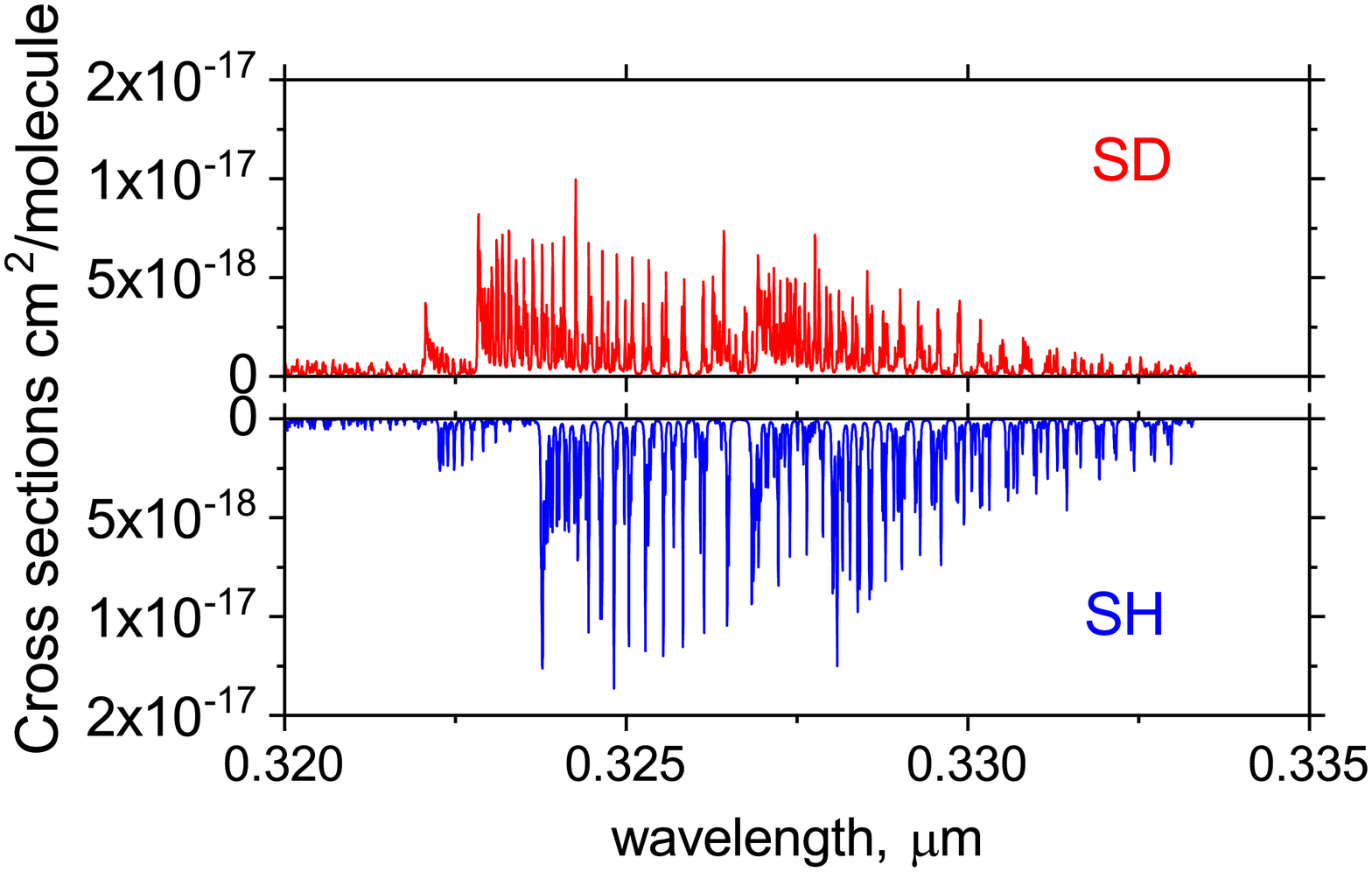}
\caption{Comparison of the $A$--$X$ (0, 0) absorption spectra ($T = 1750$~K) of SH and SD generated using our line list and the Voigt line profile with $\gamma = 0.5$~\cm\ and assuming no abundance factors.}
\label{fig:SH_SD}
\end{figure}

In order to compare our line list to experimental spectrum, we have
used \textsc{ExoCross}  to simulate spectra from our new
extended model. In Figures  \ref{fig:10TsLixx}, \ref{fig:08BuMaMo.comp},
\ref{fig:97WhOrA1-comp}, \ref{fig:88LoTixx} we
show comparisons between simulated and experimental spectra.
Figure~\ref{fig:SH_SD} compares the $A$--$X$ (0,0) bands of SH and SD
at $T = 1750$~K.
%%%%%%%%%%%%%%%%%%%%%%%%%%%%%%%%%%%%%%%%%%%%%%%%%%%%%%%%%%%%%%%%%%%%

%%%%%%%%%%%%%
%%%%%%%%%%%%%

%%%%%%%%%%%%%
%%%%%%%%%%%%%

%%%%%%%%%%%%%
%FIG. 1. Experimental ͑bottom͒ and simulated ͑top͒ spectra of part of the SH A 2⌺ ϩ – X 2⌸ ͑1,0͒ band, with rotational line assignments indicated by the combs above the spectra. The simulation was performed using spectroscopic constants from Refs. 9 and 38 and assumes a temperature of 300 K. The line shapes in the simulation are a convolution of a Gaussian ͑FWHM 0.09 cmϪ1͒ and a Lorentzian ͑FWHM 1.0 cmϪ1͒ function.
%%%%%%%%%%%%%%%%%%%%%%%%%%%%%%%%%%%%%%%%%%%%%%%%%%%%%%%%%%%%%%%%%%%%%

%%%%%%%%%%%%%
%%%%%%%%%%%%%

%%%%%%%%%%%%%
%%%%%%%%%%%%%

%%%%%%%%%%%%%%%%%%%%%%%%%%%%%%%%%%%%%%%%%%%%%%%%%%%%%%%%%%%%%%%%%%%%%
\section{Lifetimes}

The lifetime of state $i$, $\tau_i$, can be computed by summing
over the $A$ coefficients \citep{jt624}:
\begin{equation}
\tau_i =  \frac{1}{\sum_f A_{if}}~.
\end{equation}
 Our values of the lifetimes for the $v=0$ and $v=1$ vibrational states of $^{32}$SH are 449~ns and 513~ns, respectively.
The lifetimes of the $A$ states are strongly affected by the
predissocitive, forbidden interactions with (at least) one of the three
crossing states. Predissociation is not included into our current model
and therefore we do not expect the observed lifetimes to be shorter than
the ones we compute by simply considering emission lines.

The natural lifetime of the \Astate\ $v = 0$ state has been
experimentally measured to be in the range 0.5 -- 2.0 ns by
\citet{83TiFeWa.SH}, and (3 $\pm$ 2) ns by \citet{83FrBrAn.SH}.
\citet{83UbTeDy.SH} determined that the natural lifetime decreased
monotonically from (3.2 $\pm$ 0.3) ns for $N'$ = 0 to (0.95 $\pm$
0.02) ns for $N'= 9$. \citet{83FrBrAn.SH} also measured the radiative
lifetime to be (820 $\pm$ 240) ns which they note is comparable to the
\ai\ radiative lifetime calculated by \citet{85SeWeRo.SH} of 704 ns.
Later, \citet{88LoTixx.SH} used the Hanle effect and produced an
estimate of 0.17 -- 0.30 ns for the \Astate\ $v = 0$ state: this
discrepancy with previous estimates was commented on by
\citet{90UbTexx.SH} before \citet{90LoTixx.SH} provided an explanation
in terms of saturation of the transition due to power broadening using
a pulsed laser.

The predissociative lifetime of the \Astate\ $v =1$ state was
measured using cavity ring-down spectroscopy experiments by
\citet{97WhOrAsa.SH,97WhOrAs.SH} who determined values between
4.08~ps and 5.45 ps: the authors note that the \Astate\ is sensitive to
coupling with nearby dissociative $1^{4}\Sigma^{-}$,
$1^{2}\Sigma^{-}$, $1^{4}\Pi$ curves and provide a comprehensive
explanation of why their estimates are $\sim$ 5 orders of magnitude
smaller than complementary measurements of natural and radiative
lifetime.

Using our vibrationally averaged transition dipole moment $\bar\mu_{00} = \bra{X} \mu \ket{A}$ = 0.340~D we obtain the corresponding (0,0) $A$--$X$ Einstein coefficient $A_{00} = 1.026\times 10^{-6}$ 1/s and oscillator strength  $f_{00} = 0.0017$, where the following definition of the vibrational Einstein A coefficient was used:
\begin{equation}
\label{eq.vib-einsteinA}
A_{fi}^{(J=0)} = \frac{64\times 10^{-36} \pi^4 \left(\nu_{00}\right)^3}{3h} \bar\mu_{00}^2.
\end{equation}
This can be compared to the oscillator strength values $f_{00} = 0.0022$ by \citet{02BeLixx.SH} (solar spectrum) and $f_{00} = 0.0029$ by \citet{01ReOrxxa.SH} (\ai).

%by \citet{02BeLixx.SH} and $f_{00} = 0.0029$ by \citet{01ReOrx1}.

%\red{CHECK: For some reason our intensities are ~50 time weaker comparing with \citet{09ZaMaFe.SH}, who used A(0,0) from \citet{02BeLixx.SH} obtained using solar SH opacities. Our dipole however agrees well with the dipole by \citet{01ReOrx1.SH}, whose A(0,0) agrees with that by \citet{02BeLixx.SH}. I don't know what is going on but we could probably submit as it is now }

%A:=8e-36*evalf(Pi)^4*(30480.853627)^3/planck/3*(3.39952342E-01)^2;
%Using our vibronic transition moment $\bar\mu_{00} = \bra{X} \mu \ket{A}$ = 0.34~D we obtain the corresponding (0,0) $A$--$X$ Einstein coefficient $A_{00} = 0.1283\times 10^{-6}$ 1/s and oscillator strength  $f_{00} = 0.0002$.
%by \citet{02BeLixx.SH} and $f_{00} = 0.0029$ by \citet{01ReOrx1}.

The predissociation of the \Astate\ states is mainly attributed to the interaction with $1^{4}\Sigma^{-}$.
The \Astate\ state vibrational levels of the SD molecule, however are less affected by this interaction~\citep{97WhOrAs.SH}. The experimental,
collision-free lifetime reported by \citet{83TiFeWa.SH} ($v = 0$) is 189 ns. The \Duo\ value is 438~ns.
%\section{Partition function}

%%%%%%%%%%%%%%%%%%%%%%%%%%%%%%%%%%%%%%%%%%%%%%%%%%%%%%%%%%%%%%%%%%%%%%
%Updated Friday 5th July: Model54 for SH and model 69 for SD.

\begin{table}
\centering
\caption{Summary statistics for the previous SNaSH \protect{\citep{jt725}} and the new GYT line lists for SH. }
\begin{tabular}{lr|rrrrr} \hline\hline
\multicolumn{2}{c|}{\multirow{2}{*}{Statistics}}	&\multirow{2}{*}{$^{32}$SH} 	&\multirow{2}{*}{$^{33}$SH} 	& \multirow{2}{*}{$^{34}$SH} 	& \multirow{2}{*}{$^{36}$SH} & \multirow{2}{*}{$^{32}$SD} 	\\
\multicolumn{2}{c|}{}					&				&				& 				&			     &				\\ \hline	
\multirow{2}{*}{SNaSH}	& \multicolumn{1}{|c|}{Number of energies} 	& 2326	 & 2326	  & 2328 & 2334		& 4532		\\
& \multicolumn{1}{|c|}{Number of transitions}	& 81~348 	& 81~274	& 81~319 	& 81~664	& 219~463	\\ \hline
\multirow{2}{*}{GYT}		& \multicolumn{1}{|c|}{Number of energies}  &7686     &  7695 &7698 & 7709 & 12942\\
& \multicolumn{1}{|c|}{Number of transitions} & 572~145 & 573~299 & 573~639 & 575~117 & 1~127~044 \\
\hline
\hline
\end{tabular}
\label{t:stats}
\end{table}

Using the ExoMol format described by \citet{jt631}, a sample of the GYT states file is shown in Table \ref{t:SHstates} and a sample of the GYT transitions file is shown in Table \ref{t:SHtrans}.
%To effectively remove lines involving the  spurious \B\ states from any applications, the corresponding statistical weights $g_i$ in the GYT states files are set to zero.

\begin{table*}
\centering
\caption{Extract from the states file of the $^{32}$S$^{1}$H line list.   }
%Updated Friday 5th July, model 54
\tt
\label{t:SHstates}
{\tt  \begin{tabular}{rrrrrrcclrrrr} \hline \hline
$i$ & Energy (\cm) & $g_i$ & $J$ & $\tau$ & $g$-factor	& Parity & e/f	& State	& $v$	&${\Lambda}$ &	${\Sigma}$ & $\Omega$ \\ \hline
     222 &  32632.92124 &       8 &    1.5 &  5.1332E-07  &   0.66740 &   -   &   e   &  A2Sigma+  &      1 &   0   &  -0.5 &  -0.5 \\
     223 &  32720.90855 &       8 &    1.5 &  8.1317E-02  &  -0.01413 &   -   &   e   &  X2Pi      &     28 &   -1  &   0.5 &  -0.5 \\
     224 &  32943.51585 &       8 &    1.5 &  1.0814E-01  &   0.81477 &   -   &   e   &  X2Pi      &     29 &   -1  &  -0.5 &  -1.5 \\
     225 &  33233.90092 &       8 &    1.5 &  1.0856E-01  &  -0.01440 &   -   &   e   &  X2Pi      &     29 &   -1  &   0.5 &  -0.5 \\
     226 &  33482.05087 &       8 &    1.5 &  1.4501E-01  &   0.81503 &   -   &   e   &  X2Pi      &     30 &   -1  &  -0.5 &  -1.5 \\
     227 &  33773.82578 &       8 &    1.5 &  1.4573E-01  &  -0.01467 &   -   &   e   &  X2Pi      &     30 &   -1  &   0.5 &  -0.5 \\
     228 &  34046.79857 &       8 &    1.5 &  1.6400E-01  &   0.81527 &   -   &   e   &  X2Pi      &     31 &   -1  &  -0.5 &  -1.5 \\
     229 &  34221.19840 &       8 &    1.5 &  5.8938E-07  &   0.66739 &   -   &   e   &  A2Sigma+  &      2 &   0   &  -0.5 &  -0.5 \\
     230 &  34339.87891 &       8 &    1.5 &  1.6466E-01  &  -0.01492 &   -   &   e   &  X2Pi      &     31 &   -1  &   0.5 &  -0.5 \\
     231 &  34637.04539 &       8 &    1.5 &  1.5672E-01  &   0.81550 &   -   &   e   &  X2Pi      &     32 &   -1  &  -0.5 &  -1.5 \\
     232 &  34931.35680 &       8 &    1.5 &  1.5718E-01  &  -0.01516 &   -   &   e   &  X2Pi      &     32 &   -1  &   0.5 &  -0.5 \\
     233 &  35252.16294 &       8 &    1.5 &  1.4828E-01  &   0.81573 &   -   &   e   &  X2Pi      &     33 &   -1  &  -0.5 &  -1.5 \\
     234 &  35547.63821 &       8 &    1.5 &  1.4862E-01  &  -0.01539 &   -   &   e   &  X2Pi      &     33 &   -1  &   0.5 &  -0.5 \\
     235 &  35601.52344 &       8 &    1.5 &  6.8906E-07  &   0.66736 &   -   &   e   &  A2Sigma+  &      3 &   0   &  -0.5 &  -0.5 \\
     236 &  35891.59274 &       8 &    1.5 &  1.5436E-01  &   0.81595 &   -   &   e   &  X2Pi      &     34 &   -1  &  -0.5 &  -1.5 \\
     237 &  36188.17116 &       8 &    1.5 &  1.5475E-01  &  -0.01561 &   -   &   e   &  X2Pi      &     34 &   -1  &   0.5 &  -0.5 \\
     238 &  36554.83499 &       8 &    1.5 &  1.8246E-01  &   0.81617 &   -   &   e   &  X2Pi      &     35 &   -1  &  -0.5 &  -1.5 \\
     239 &  36758.86698 &       8 &    1.5 &  8.3812E-07  &   0.66729 &   -   &   e   &  A2Sigma+  &      4 &   0   &  -0.5 &  -0.5 \\
     240 &  36852.46455 &       8 &    1.5 &  1.8357E-01  &  -0.01583 &   -   &   e   &  X2Pi      &     35 &   -1  &   0.5 &  -0.5 \\
     241 &  37241.44004 &       8 &    1.5 &  2.4394E-01  &   0.81638 &   -   &   e   &  X2Pi      &     36 &   -1  &  -0.5 &  -1.5 \\
     242 &  37540.07857 &       8 &    1.5 &  2.4736E-01  &  -0.01604 &   -   &   e   &  X2Pi      &     36 &   -1  &   0.5 &  -0.5 \\
     243 &    46.129449 &      12 &    2.5 &  1.9375E+02  &   0.38573 &   +   &   e   &  X2Pi      &      0 &   1   &   0.5 &   1.5 \\
     244 &   445.285462 &      12 &    2.5 &  4.0720E+01  &  -0.04273 &   +   &   e   &  X2Pi      &      0 &   1   &  -0.5 &   0.5 \\
     245 &  2642.838347 &      12 &    2.5 &  6.8587E-01  &   0.38456 &   +   &   e   &  X2Pi      &      1 &   1   &   0.5 &   1.5 \\
     246 &  3041.672932 &      12 &    2.5 &  6.6743E-01  &  -0.04156 &   +   &   e   &  X2Pi      &      1 &   1   &  -0.5 &   0.5 \\
\hline
\hline
%%%%
%%%%
\end{tabular}}
\mbox{}\\
{\flushleft
$i$:   State counting number.     \\
$\tilde{E}$: State energy in \cm. \\
$g_i$:  Total statistical weight, equal to ${g_{\rm ns}(2J + 1)}$.     \\
$J$: Total angular momentum.\\
$\tau$: Lifetime (s$^{-1}$).\\
$g$: Land\'{e} $g$-factors. \\
$+/-$:   Total parity. \\
$e/f$:   Rotationless parity. \\
State: Electronic state.\\
$v$:   State vibrational quantum number. \\
$\Lambda$:  Projection of the electronic angular momentum. \\
$\Sigma$:   Projection of the electronic spin. \\
$\Omega$:   Projection of the total angular momentum, $\Omega=\Lambda+\Sigma$. \\
}
\end{table*}
%%%%%%%%%%%%%%%%%%%%%%%%%%%%%%%%%%%%%%%%%%%%%%%%%%%%%%%%%%%%%%%%%%%%%%%%%%%%%%%%%%%%%%%%%%%%%%%%%%%%%%%%%%%%%%%%%%%%%%%%%%%%%%%%%%%%%%%%%%%%

%%%%%%%%%%%%%%%%%%%%%%%%%%%%%%%%%%%%%%%%%%%%%%%%%%%%%%%%%%%%%%%%%%%%%%%%%%%%%%%%%%%%%%%%%%%%%%%%%%%%%%%%%%%%%%%%%%%%%%%%%%%%%%%%%%%%%%%%%%%%
\begin{table}
\centering
\caption{Extract from the transitions file of the $^{32}$SH line list.}
\tt
\label{t:SHtrans}
\centering
\begin{tabular}{rrrr} \hline\hline
\multicolumn{1}{c}{$f$}	&	\multicolumn{1}{c}{$i$}	& \multicolumn{1}{c}{$A_{fi}$ (s$^{-1}$)}	&\multicolumn{1}{c}{$\tilde{\nu}_{fi}$} \\ \hline
        3438    &     3157 & 6.6514E-05    &    30001.408693  \\
        1314    &     1188 & 1.7677E-07    &    30002.099401  \\
         365    &       85 & 4.7787E-07    &    30002.250973  \\
         286    &      164 & 4.5799E-07    &    30002.276822  \\
         716    &      407 & 5.7148E-07    &    30002.319009  \\
         637    &      486 & 4.8517E-06    &    30002.383604  \\
        2110    &     1966 & 1.4836E+05    &    30002.561088  \\
        2396    &     2418 & 1.0166E-05    &    30002.770705  \\
        2859    &     2866 & 2.3905E-05    &    30003.312455  \\
        1781    &     1655 & 4.3074E-08    &    30003.351508  \\
        1729    &     1891 & 4.0626E-07    &    30003.584108  \\
        1704    &     1732 & 1.1467E-10    &    30004.141218  \\
         380    &      403 & 1.4745E+05    &    30004.344227  \\
        1803    &     1659 & 1.4449E-09    &    30004.767420  \\
        3367    &     3229 & 6.6557E-05    &    30005.009831  \\
         708    &      406 & 4.9356E-06    &    30005.192773  \\
         538    &      403 & 2.0426E+05    &    30005.269387  \\
\hline\hline
\end{tabular} \\ \vspace{2mm}
\rm
\noindent
$f$: Upper  state counting number;\\
$i$:  Lower  state counting number; \\
$A_{fi}$:  Einstein-$A$ coefficient in s$^{-1}$; \\
$\tilde{\nu}_{fi}$: transition wavenumber in \cm.\\
\end{table}
%%%%%%%%%%%%%%%%%%%%%%%%%%%%%%%%%%%%%%%%%%%%%%%%%%%%%%%%%%%%%%%%%%%%%%%%%%%%%%%%%%%%%%%%%%%%%%%%%%%%%%%%%%%%%%%%%%%%%%%%%%%%%%%%%%%%%%%%%%%%

%%%%%%%%%%%%%%%%%%%%%%%%%%%%%%%%%%%%%%%%%%%%%%%%%%%%%%%%%%%%%%%%%%%%%%%%%%%%%%%%%%%%%%%%%%%%%%%%%%%%%%%%
\section{Conclusions}

We extend the previous ExoMol line list which covered only the \X\ state of
the mercapto (SH) to include transitions within the \Astate - \X\ system.
This new experimentally-tuned theoretical line list, GYT, supersedes the previous ExoMol SNaSH line list \citep{jt725} and the
existing experimental, limited, absorption-spectrum line list of
\citet{09ZaMaFr.SH} which has previously been successfully used to
model Hot Jupiters. Using available experimental data we have
generated a model for SH and a separate model for SD. These new line
lists now extend out into the UV regime (to $\sim$ 0.256 $\mu$m) and
cover transitions up to $\sim$ 39~000 \cm.

As with all line lists produced by the ExoMol the validity of models
is dependent on the vibrational and, to a lesser extent, rotational
ranges of experimental data: here we have been limited to $v' \leq 2$.
Using the custom-built \Duo\ programme, an array of coupling
contributions have been included to account for nearby dissociative
electronic states and higher-order breakdown terms. The accuracy of
models is also dependent on the underlying measurement uncertainties
of available experimental data: here high-level \ai\ curves have been
refined to accuracies similar to the available experimental data
($\sim$0.3 \cm). These line lists, are available from the CDS
\href{http://cdsarc.u-strasbg.fr}{http://cdsarc.u-strasbg.fr} and
ExoMol \href{www.exomol.com}{www.exomol.com} data bases.
States files are provided as part of the supplementary material to this paper together with the spectroscopic models
in the form of the \Duo\ input files.

The spectral range and accuracy of our line lists taken together
corresponds to a resolving power $\sim$ 60~000 hence making these line
list of use within the developing research niche of using high
resolution planetary radial velocity measurements to characterise an
exoplanetary atmospheres. At present the ExoMol group has published
several line lists which, owing to required experimental data been
available, could be of use for this method including most recently TiO
\citep{jt760}. In order to produce such a line list of this desired
accuracy, the MARVEL process \citep{jt412} can be used to generate
what can be regarded as an ``experimental'' list of rovibronic
energies from measured transitions. This process has already been used
for various molecules of interest in exoplanets and cool stellar
objects including ZrO \citep{jt740}, TiO \citep{jt672}, C$_{2}$
\citep{jt637}, C$_{2}$H$_{2}$ \citep{jt705}, H$_{2}$S \citep{jt718},
NH$_{3}$ \citep{jt608} and most recently NH
\citep{jt764}.%Note MARVEL not used here - trying to make sep. point about ExoMol moving to sep out accuracy (RV method) and completeness (atmospheric).

Our new extended line list has the benefit of having the required \textit{accuracy} for use in high-resolution spectroscopy and the relative property of \textit{completeness} (up to $\sim$ 5000 K) owing to its extended spectral range. We envisage this will help benefit the characterisation of hot Jupiter exoplanets such as WASP-121b in which SH is suspected to contribute to opacity within the UV regime.
%Comment about MARVEL process?
%Previous SH paper - listed out SH molecules.
%%%%%%%%%%%%%%%%%%%%%%%%%%%%%%%%%%%%%%%%%%%%%%%%%%%%%%%%%%%%%%%%%%%%%%%%%%%%%%%%%%%%%%%%%%%%%%%%%%%%%%%%%

%%%%%%%%%%%%%%%%%%%%%%%%%%%%%%%%%%%%%%%%%%%%%%%%%%%%%%%%%%%%%%%%%%%%%%%%%%%%%%%%%%%%%%%%%%%%%%%%%%%%%%%%%
\section{Acknowledgements}
This work was supported by the UK Science and Technology Research
Council (STFC) No. ST/R000476/1.
This work made extensive use of UCL's Legion high performance
computing facility  along with the STFC DiRAC HPC facility
supported by BIS National E-infrastructure capital grant ST/J005673/1
and STFC grants ST/H008586/1 and ST/K00333X/1. Some support was provided by the NASA Laboratory Astrophysics program.
%%%%%%%%%%%%%%%%%%%%%%%%%%%%%%%%%%%%%%%%%%%%%%%%%%%%%%%%%%%%%%%%%%%%%%%%%%%%%%%%%%%%%%%%%%%%%%%%%%%%%%%%%

%%%%%%%%%%%%%%%%%%%%%%%%%%%%%%%%%%%%%%%%%%%%%%%%%%%%%%%%%%%%%%%%%%%%%%%%%%%%%%%%%%%%%%%%%%%%%%%%%%%%%%%%%
%\bibliographystyle{mnras}
%\bibliography{journals_astro,SH,programs,methods,linelists,jtj,partition,exoplanets,exogen,SiH,PS,AlO,abinitio}
%\label{lastpage}
\end{document}